\documentstyle[12pt,epsfig]{article}
\renewcommand{\thesection}{\Roman{section}.}

\newcommand{\Frac}[2]{\frac{\displaystyle #1}{\displaystyle #2}}
\newcommand{\kpg}{$K^+ \rightarrow \pi^+ \gamma \gamma$}
\newcommand{\kpign}{$K_L \rightarrow \pi^0 \gamma \gamma$}
\newcommand{\kpgee}{$K^+ \rightarrow \pi^+ \gamma e^+ e^-$}
\newcommand{\kpgmm}{$K^+ \rightarrow \pi^+ \gamma \mu^+ \mu^-$}
\newcommand{\n}{\noindent}
\newcommand{\no}{\nonumber}
\def\ap#1#2#3{{Ann. Phys. (N.Y.) {\bf #1}, #2 (#3)}}
\def\ijm#1#2#3{{Int. J. Mod. Phys. A {\bf #1}, #2 (#3)}}
\def\nc#1#2#3{{Nuovo Cimento {\bf A~#1}, #2 (#3)}}
\def\npb#1#2#3{{Nucl. Phys. {\bf B#1}, #2 (#3)}}
\def\plb#1#2#3{{Phys. Lett. B {\bf #1}, #2 (#3)}}
\def\prr#1#2#3{{Phys. Rev. {\bf #1}, #2 (#3)}}
\def\rmp#1#2#3{{Rev. Mod. Phys. {\bf #1}, #2 (#3)}}
\def\prd#1#2#3{{Phys. Rev. D {\bf #1}, #2 (#3)}}
\def\prl#1#2#3{{Phys. Rev. Lett. {\bf #1}, #2 (#3)}}

\def\ptp#1#2#3{{Prog. Theor. Phys. {\bf #1}, #2 (#3)}}
\def\zp#1#2#3{{Z. Phys. C {\bf #1}, #2 (#3)}}
\def\epj#1#2#3{{Eur. Phys. J. C {\bf #1}, #2 (#3)}}
\def\jep#1#2#3{{J. High Energy Phys. {\bf #1}, #2 (#3)}}

\def\pr0{p^{\rho}_0}
\def\ps0{p^{\sigma}_0}
\def\p0m{p^{\phantom{l}}_{0\mu}}
\def\pn0{p^{\phantom{l}}_{0\nu}}
\def\ri{\rightarrow}

\def\gsim{\ \rlap{\raise 2pt \hbox{$>$}}{\lower 2pt \hbox{$\sim$}}\ }
\def\lsim{\ \rlap{\raise 2pt \hbox{$<$}}{\lower 2pt \hbox{$\sim$}}\ }

\def\a2{{8 \over {3 r^4_{\pi}}}{(4 \zeta_1+\xi_1)}}
\def\as{\arctan{\sqrt{s \over {4 m^2_{\pi}-s}}}}
\def\ak{\arctan{\sqrt{k^2_1 \over {4 m^2_{\pi}-k^2_1}}}}
\def\sa{\sqrt{4 m^2_{\pi}-s}}
\def\ka{\sqrt{4 m^2_{\pi}-k^2_1}}
\def\ds{(s-k^2_1)}
\def\ns{(8 m^4_{\pi}+2 m^2_{\pi} s -s^2)}
\def\nk{(8 m^4_{\pi}+2 m^2_{\pi} k^2_1 -k^4_1)}
\def\mp{m_{\pi}}
\def\k1{k_1}
\def\nns{(8 m^4_{\pi}-6 m^2_{\pi} s +s^2)}
\def\nnk{(8 m^4_{\pi}-6 m^2_{\pi} k^2_1 +k^4_1)}
\begin{document}

\sloppy
\vskip -2.0truecm
{\hfill DUKE-TH-98-175}

\bigskip\bigskip\medskip

\centerline{\Large {\bf Unitarity corrections to \mbox{\boldmath $K^+
\rightarrow \pi^+ \gamma l^+ l^-$}}}

\medskip
\bigskip\bigskip\medskip

\centerline{\large{Fabrizio Gabbiani}}
\medskip

\centerline{\large{Department of Physics}}
\smallskip

\centerline{\large{Duke University, Durham, North Carolina 27708}}

\vskip 1.0 truecm

\begin{abstract}

We perform a chiral one-loop calculation of the unitarity corrections
to the processes $K^+ \rightarrow \pi^+ \gamma l^+ l^-$ up to $\cal
O$(E$^6)$, taking into account $\pi^+\pi^-$ intermediate states. Branching
ratios and differential branching ratios are computed and presented to
demonstrate the importance of the above corrections.

\end{abstract}


\section{Introduction}

The investigation of radiative rare kaon decays has taught us that
they form a complex of interrelated
processes which share some common features. Experimental and
theoretical results on any of these reactions are useful in the
analysis of all of them. Since they can be analyzed using
chiral perturbation theory (ChPT) \cite{CPT}, the experimental
exploration of the entire complex provides stringent checks on this
theoretical method. Recently, radiative kaon decays have attracted
considerable attention from the theoretical \cite{DEIN,DG1,DG2,DEIP}
and the experimental \cite{KTEV,KTEV1} communities. Predictions
for many of them already exist in the literature
\cite{EPR,CEP,CDM,DG,DP,DP1,VMD}, and several new experimental
investigations are under way or planned. In particular, initial data
on the process $K_L \ri \pi^0 \gamma e^+ e^-$ \cite{DG1} have
already been presented \cite{KTEV1}, together with further data on
\kpign. Confident that the objects of our calculation, the decays $K^+
\ri \pi^+ \gamma l^+ l^-$, are accessible to experiment, our
goal is to provide information on their rate and the corresponding
decay distributions.

Analogously to what has been studied in the case of \kpign, the
reaction \kpg\ takes place predominantly through loop diagrams with
pions in the loop. In the former process, the decay distribution is
quite distinctive and the rate is predicted without any free
parameters at one-loop order. While the distribution agrees well with
experiment, the theoretical rate appears too small by more than a
factor of 2. Because of this, several authors have gone beyond the
straightforward one-loop (order E$^4$) chiral calculation. Adding a
series of higher order effects in a quasi-dispersive framework one has
a surprising success at increasing the rate without modifying the
decay distribution greatly \cite{CDM,CEP}. The process \kpg\ has been
studied in a similar way \cite{DP,DP1}. The physics which determines
the above decays is also involved in the reaction we consider. The
experimental study of the lepton-photon modes can achieve independent
confirmation of the dynamics that drives the whole complex of decay
modes.

As was done in previous works, we separately calculate the one-loop
results within ChPT both at ${\cal O}$(E$^4$) and ${\cal O}$(E$^6$).
The first one gives a prediction for the rate and the variation of the
amplitude depending on the invariant mass of the two leptons, $k^2_1$,
which is carried by an off-shell photon. An additional parameter, in
the form of a local counterterm, also enters the calculation. In the
second case, following Ref. \cite{DP}, we take into account the higher
order behavior in the experimental $K^+ \ri 3 \pi$ decay rate.

At ${\cal O}$(E$^6$) the higher order effects in the $K^+ \ri 3 \pi$
vertex are extracted from a quadratic fit to the amplitude. According
to the results of Refs. \cite{DP,DP1}, we shall not be concerned with
vector meson corrections, likely too small to be significant, given
the uncertainties in the several parameters involved in this
computation.
\vfill\eject

This paper is organized as follows: In Sec. II we fix our notation and
define the quantities used in the rest of this paper by summarizing
some established results for the decay \kpg. This provides a starting
point for our calculation, taking a photon off shell and going through
the process $K^+$ $\ri$ $\pi^+ \gamma \gamma^*$ $\ri$ $\pi^+ \gamma
l^+ l^-$. In Sec. III we describe the ${\cal O}$(E$^4$)
calculation, which we extend to ${\cal O}$(E$^6$) in Sec. IV, taking
into account the unitarity corrections at one loop. Finally, we
recapitulate our conclusions in Sec. V. All the relevant expressions
for the integrals used in this paper are shown in the Appendix.

\section{\mbox{\boldmath $K \ri \pi \gamma \gamma$ amplitudes}}

Let us first review some previously known results for $K \ri
\pi \gamma \gamma$, and establish our notation for the following
sections. We define the general amplitude for $K \ri \pi
\gamma \gamma$ as given by

\begin{equation}
M[K (p^{\phantom{l}}_K) \ri \pi (p) \gamma
(k_1,\epsilon_1)
\gamma (k_2,\epsilon_2) ] = {\epsilon_1}_{\mu} {\epsilon_2}_{\nu}
 {\cal M}^{\mu \nu} (p^{\phantom{l}}_K,k_1,k_2)
\end{equation}

\n where $\epsilon_1$,$\epsilon_2$ are the photon polarizations, and
${\cal M}^{\mu \nu}$ has four invariant amplitudes:

\begin{eqnarray}
{\cal M}^{\mu \nu} & = & A(z,y) ( k_2^{\mu} k_1^{\nu} -
k_1 \cdot k_2 g^{\mu \nu}) \no \\
&+& B(z,y) \left({{p^{\phantom{l}}_K \cdot k_1 p \cdot
k_2} \over {k_1 \cdot k_2}} g^{\mu \nu}
+ p^{\mu}_K p^{\nu}_K
- {{p_K \cdot k_1} \over {k_1 \cdot k_2}} k_2^{\mu} p^{\nu}_K
- {{p_K \cdot k_2} \over {k_1 \cdot k_2}} p^{\mu}_K k_1^{\nu} \right)
\no \\
& + & C_1(z,y) \varepsilon^{\mu \nu \rho \sigma}
{k_1}_{\rho} {k_2}_{\sigma} \no \\
&+& C_2(z,y) \left[\varepsilon^{\mu \nu \rho \sigma}
 {{p^{\phantom{l}}_K \cdot k_2 {k_1}_{\rho} + p^{\phantom{l}}_K
\cdot k_1 {k_2}_{\rho}} \over {k_1 \cdot k_2}}p^{\phantom{l}}_{K\sigma}
+(p^{\mu}_K \varepsilon^{\nu \alpha \beta \gamma} + p^{\nu}_K
\varepsilon^{\mu \alpha \beta \gamma} ) p_{K\alpha} {{{k_1}_{\beta}
{k_2}_{\gamma}} \over {k_1 \cdot k_2}} \right] \label{tensor}
\end{eqnarray}

\n where

\begin{equation}
y = \Frac{p^{\phantom{l}}_K \cdot (k_1 - k_2)}{m_K^2},
\qquad\qquad z = \Frac{(k_1 + k_2)^2}{m_K^2}.
\end{equation}

\n The physical region in the adimensional variables $y$ and $z$ is
given by

\begin{equation}
0 \leq |y| \leq \Frac{1}{2} \lambda^{1/2}(1,r_{\pi}^2,z),
\qquad\qquad 0 \leq z \leq (1-r_{\pi})^2,
\end{equation}

\n with

\begin{equation}
\lambda(1, z, r^2 ) = 1 + z^2 + r^4 - 2z - 2r^2 - 2r^2z, \qquad
r_{\pi} = \Frac{m_{\pi}}{m_K}. \label{lam} \\
\end{equation}

\n Here $k_1$ and $k_2$ are the momenta of the off-shell and on-shell
photons, respectively, with the off-shell photon materializing into
the lepton pair. Note that the invariant amplitudes $A(z,y)$, $B(z,y)$
and $C_1(z,y)$ have to be symmetric under the interchange of $k_1$ and
$k_2$ as required by Bose symmetry, while $C_2(z,y)$ is antisymmetric.

Using the definitions (\ref{tensor})--(\ref{lam}) the double differential
rate for unpolarized photons is given by

\begin{eqnarray}
\Frac{d^2 \Gamma}{d y d z} & = &
\Frac{m^5_K}{2^9 \pi^3} \left\{ z^2 \left( \left|A - {B
\over 2}\right|^2 +|C_1|^2 \right)\right. \no \\
&+& \left. \left[y^2 - \Frac{1}{4} \lambda (1,r_{\pi}^2,z) \right]^2
\left({{|B|^2} \over 4} + |C_2|^2 \right) \right\}.
\end{eqnarray}

\n In the limit where $CP$ is conserved, the amplitudes $A$ and $B$
contribute to $K_2\ri\pi^0\gamma\gamma$ whereas
$K_1\ri\pi^0\gamma\gamma$ involves the other two amplitudes $C_1$ and
$C_2$. All four amplitudes contribute to \kpg.
Only $A$ and $C_1$ are non-vanishing to lowest non-trivial order, ${\cal
O}$(E$^4)$, in ChPT. As argued in \cite{DI1,DP}, the antisymmetric
character of the $C_2(z,y)$ amplitude under the interchange of $k_1$
and $k_2$ means effectively that while its leading contribution is
${\cal O}$(E$^6)$, this can only come from a finite loop calculation
because the leading counterterms for the $C_2$ amplitude are
${\cal O}$(E$^8)$. Moreover, this loop contribution is helicity suppressed
compared to the $B$ term. This antisymmetric ${\cal O}$(E$^6)$ loop
contribution might be smaller than the local ${\cal O}$(E$^8)$ contribution.

\section{\mbox{\boldmath $\cal{O}$(E$^4$) calculation}}

First let us provide the straightforward $\cal{O}$(E$^4$) calculation
of ${\cal M}(K^+ \ri \pi^+ \gamma l^+ l^-)$ within ChPT. This
is the generalization to $k^2_1 \neq 0$ of the original chiral
calculation of the authors of \cite{EPR1,EPR2}, and it includes all the
$k^2_1/m^2_{\pi}$ and  $q$ $\equiv$ $k^2_1/m^2_K$ variations of the
amplitudes at this order in the energy expansion. There can be further
$k^2_1$/(1 GeV)$^2$ corrections which correspond to $\cal{O}$(E$^6$)
and higher. The easiest technique for this calculation uses the basis
where the kaon and pion fields are transformed so that the propagators
have no off-diagonal terms, as described in
Refs. \cite{EPR1,EPR2}. Some of the relevant diagrams are shown in
Fig. 1.

\begin{figure}[htbp]
\centering
\leavevmode
\centerline{
\epsfbox{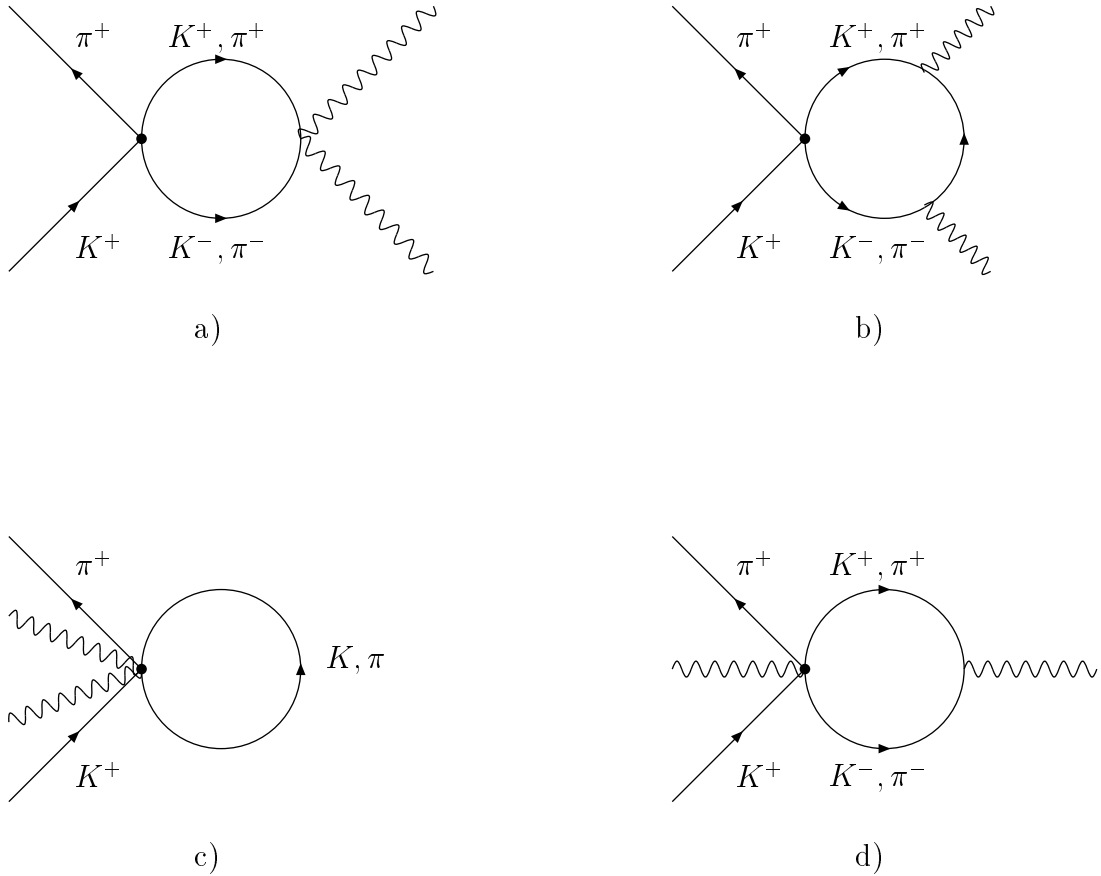}}
\caption{Some diagrams relevant to the process $K^+ \ri
\pi^+ \gamma l^+ l^-$ at ${\cal O}$(E$^4)$ and ${\cal O}$(E$^6)$.
Either photon may also be radiated from the incoming $K^+$ or
the outcoming $\pi^+$. The lepton pair must be attached
to one of the photons, and the on-shell photon may be radiated from
one of these leptons.}
\end{figure}

Analogously to the leading $\Delta I = 1/2$ ${\cal O}$(E$^4)$ $A(z,y)$ and
$C_1(z,y)$ amplitudes for \kpg\, which have been computed in
\cite{EPR2}, we can write an expression for the $A^{(4)}$ amplitude
for $K^+ \ri \pi^+ \gamma \gamma^*$,

\begin{equation}
A^{(4)}(z) = \Frac{G_8 \alpha_{em}}{2 \pi (z-q)} \left\{
 (z+1-r_{\pi}^2) [1+2 I(m^2_{\pi})] +
(z+r_{\pi}^2-1)[1+2 I(m^2_K)] - \hat{c} (z-q) \right\},
\label{a4}
\end{equation}

\n where $G_8$ is the effective weak coupling constant determined from
$K \ri \pi \pi$ decays at ${\cal O}$(E$^2)$:

\begin{eqnarray}
G_8 & = & {G_F \over \sqrt{2}} \vert V^{\phantom{*}}_{ud} V^*_{us} \vert
g^{\phantom{\dagger}}_8, \no \\
g^{\rm tree}_8 & = & 5.1,
\end{eqnarray}

\n where $V$ is the Cabibbo-Kobayashi-Maskawa matrix \cite{KM}, and

\begin{eqnarray}
I(m^2_{\pi}) &=& \int^1_0 dz_1 \int_0^{1-z_1} dz_2
{{m^2_{\pi}-z_1(1-z_1)k^2_1} \over {2 z_1 z_2 k_1 \cdot k_2
+z_1(1-z_1)k^2_1 -m^2_{\pi} +i\epsilon}} \no \\
&=& {m^2_{\pi} \over {s -k^2_1}} [F(s) - F(k^2_1)] - {k^2_1 \over {s
-k^2_1}} [G(s) - G(k^2_1)].
\end{eqnarray}

\n The notation is defined by

\begin{equation}
s = (p^{\phantom{l}}_K-p^{\phantom{l}}_+)^2=(k_1 + k_2)^2
\end{equation}

\n and

\begin{equation}
F(a) = \int^1_0 {dz_1 \over z_1} \log\left[{{m^2_{\pi} - a(1-z_1)z_1 -
i \epsilon} \over {m^2_{\pi}}}\right],
\end{equation}

\begin{equation}
G(a) = \int^1_0 dz_1 \log\left[{{m^2_{\pi} - a(1-z_1)z_1 -
i \epsilon} \over {m^2_{\pi}}}\right].
\end{equation}

\n The above functions are related to those presented in Ref.
\cite{CEP},

\begin{equation}
F(a) = {a \over {2 m^2_{\pi}}} \left[F_{\rm CEP}\left({a \over
4 m^2_{\pi}}\right)-1\right],
\end{equation}

\begin{equation}
G(a) = -{a \over {2 m^2_{\pi}}} \left[R_{\rm CEP}\left({a \over
4 m^2_{\pi}}\right)+{1 \over 6}\right],
\end{equation}

\n which are shown below:

\begin{eqnarray}
F_{\rm CEP}(x) & = & 1 - {1 \over x} \left[ \arcsin \left(
\sqrt{x} \right) \right]^2 \qquad (x \leq 1) \no \\
& = & 1 + {1 \over 4x} \left( \log {1 - \sqrt{1 - 1/x} \over 1 + \sqrt{1 -
1/x}} + i \pi \right)^2 \qquad (x \geq 1),  \\
R_{\rm CEP}(x) & = & - {1 \over 6} + {1 \over 2x} \left[ 1 - \sqrt{1/x
- 1} \arcsin \left( \sqrt{x} \right) \right]
\qquad (x \leq 1) \no \\
& & - {1 \over 6} + {1 \over 2x} \left[ 1 + \sqrt{1 - 1/x} \left( \log {1 -
\sqrt{1 - 1/x} \over 1 + \sqrt{1 - 1/x}} + i \pi \right) \right],
\qquad (x \geq 1). \no \\
\phantom{l}
\end{eqnarray}

\n The above results agree with the results obtained in
\cite{DP} in the $k^2_1 \ri 0$ limit.

In Eq. (\ref{a4}) the pion loop contribution largely dominates
over the kaon loop part. The loop results are finite, but ChPT allows
an ${\cal O}$(E$^4)$ scale independent local contribution that
may be parametrized as \cite{KMW1}

\begin{equation}
\hat{c} ={{128 \pi^2}\over{3}}[3(L_9+L_{10}) +N_{14}-N_{15}-2N_{18}]
\end{equation}

\n or, using the notation of \cite{EPR2,EPR3},

\begin{equation}
\hat{c} ={{32 \pi^2}\over{3}}[12(L_9+L_{10}) -w_1 - 2 w_2 - 2 w_4],
\end{equation}

\n where $\hat c$ is a quantity of ${\cal O}(1)$. The $L_9$ and $L_{10}$ are
the local ${\cal O}$(E$^4)$ strong couplings and $N_{14},N_{15}$ and
$N_{18}$ (or $w_1$, $w_2$ and $w_4$) are ${\cal O}$(E$^4)$ weak
couplings, still not completely fixed by the phenomenology, and which
can be only computed in a model dependent way \cite{EKW}. The weak
deformation model (WDM) \cite{EPR3} predicts $\hat{c}=0$, while naive
factorization in the factorization model (FM) \cite{PR,EKW} gives
$\hat{c} = -2.3$. In these models, because of the cancellation in the
vector meson contribution in $\hat{c}$, the role of axial mesons could
be relevant \cite{EKW}.

Thirty-one events for the process \kpg\ have been observed at BNL (E787)
\cite{K}, with the partial branching ratio
BR(\kpg, 100 MeV/$c$ $ < P_{\pi^+} < $
180 MeV/$c$) = $(6.0\pm1.5 \mbox{\{stat\}} \pm0.7
\mbox{\{syst\}})\times 10^{-7}$. This has been extrapolated with the
help of ChPT, performing a maximum likelihood fit of $\hat{c}$
to the spectrum. The results of the fit to the data support the
inclusion of the unitarity corrections, giving as the best fit
$\hat{c}$ = 1.8 $\pm$ 0.6 and BR(\kpg) =
(1.1 $\pm$ 0.3 $\pm$ 0.1) $\times$ 10$^{-6}$, as also reported in the
Review of Particle Physics \cite{RPP}.

The ${\cal O}$(E$^4)$ contribution to the $C_1(z,y)$ amplitude is

\begin{equation}
C_1(z) = \Frac{G_8 \alpha_{em}}{\pi} \left[
\Frac{z - r_{\pi}^2}{z - r_{\pi}^2 +
i r_{\pi} \Frac{\Gamma_{\pi^0}}{m_K}} -
\Frac{z - \Frac{2 + r_{\pi}^2}{3}}{z - r_{\eta}^2} \right] ,
\end{equation}

\n where $r_{\eta} = m_{\eta}/m_K$ and $\Gamma_{\pi^0} \equiv
\Gamma ( \pi^0 \ri \gamma \gamma^* )$ $\sim$ 0. This amplitude
is generated by the Wess--Zumino--Witten functional \cite{WZW}
$(\pi^0, \eta) \ri \gamma \gamma^*$ through the sequence
$K^+ \ri \pi^+ (\pi^0,\eta) \ri \pi^+ \gamma
\gamma^*$. This contribution amounts to less than $10\%$ in the
total width.

The ${\cal O}$(E$^4)$ results can be expressed as total branching
ratios. They are summarized in Table I for three values of $\hat{c}$,
given respectively by the weak deformation model, the factorization
model, and the fit to BR(\kpg) mentioned above.

\begin{table}[htbp]
\centering
\caption{Results for BR($K^+\ri\pi^+\gamma l^+l^-$) at ${\cal O}$(E$^4)$.}
\vskip 0.1 in
\begin{tabular}{ccc} \hline\hline\\ 
& BR(\kpgee) & BR(\kpgmm) \\
\hline\\
$\hat{c}$ = 1.8 (fit) & 1.4 $\times$ 10$^{-8}$ & 3.9 $\times$ 10$^{-11}$ \\
$\hat{c}$ = 0 (WDM) & 8.6 $\times$ 10$^{-9}$ & 3.6 $\times$ 10$^{-11}$ \\
$\hat{c}$ = --2.3 (FM) & 5.7 $\times$ 10$^{-9}$ & 3.9 $\times$ 10$^{-11}$ \\

\hline\hline
\end{tabular}
\end{table}

The decay distributions in $z$ and $y$ provide more detailed
information. We present them in Figs. 2--5.

\begin{figure}[htbp]
\vfill
\centerline{
\begin{minipage}[t]{.44\linewidth}\centering
\mbox{\epsfig{file=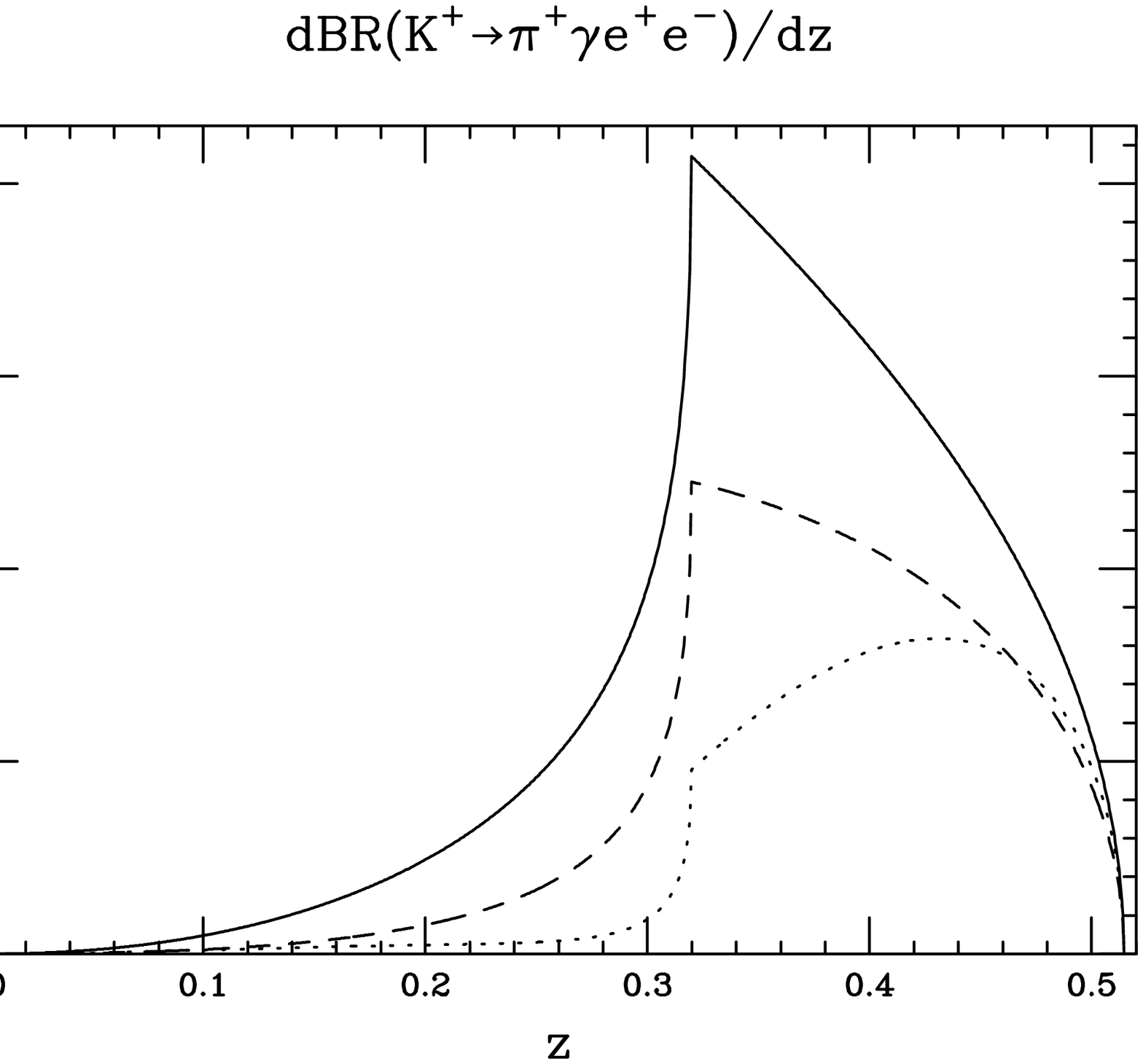,width=7.0cm}}
\caption{The differential branching ratio dBR(\kpgee)/dz
to order E$^4$ is plotted
vs $z$ for $\hat c$ = 1.8 (solid line), $\hat c$ =
0 (dashed line) and $\hat c$ = --2.3 (dotted line).}
\end{minipage}
\hspace{1.0cm}
\begin{minipage}[t]{.44\linewidth}\centering
\mbox{\epsfig{file=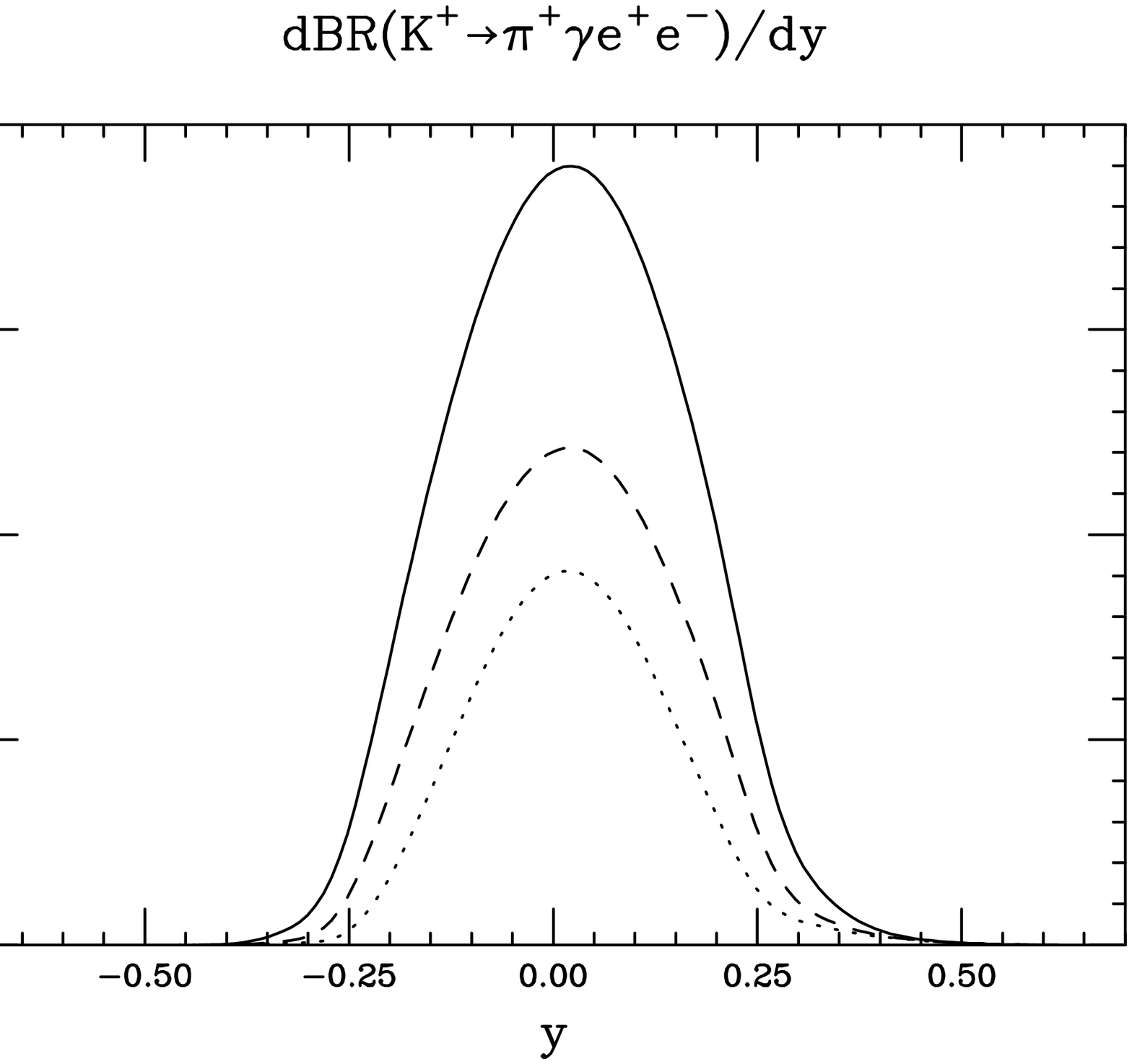,width=7.0cm}}
\caption{The differential branching ratio
dBR(\kpgee)/dy
to order E$^4$ is plotted vs $y$ for $\hat c$ = 1.8 (solid line), $\hat c$ =
0 (dashed line) and $\hat c$ = --2.3 (dotted line).}
\end{minipage}}
\end{figure}

\begin{figure}[htbp]
\vfill
\centerline{
\begin{minipage}[t]{.44\linewidth}\centering
\mbox{\epsfig{file=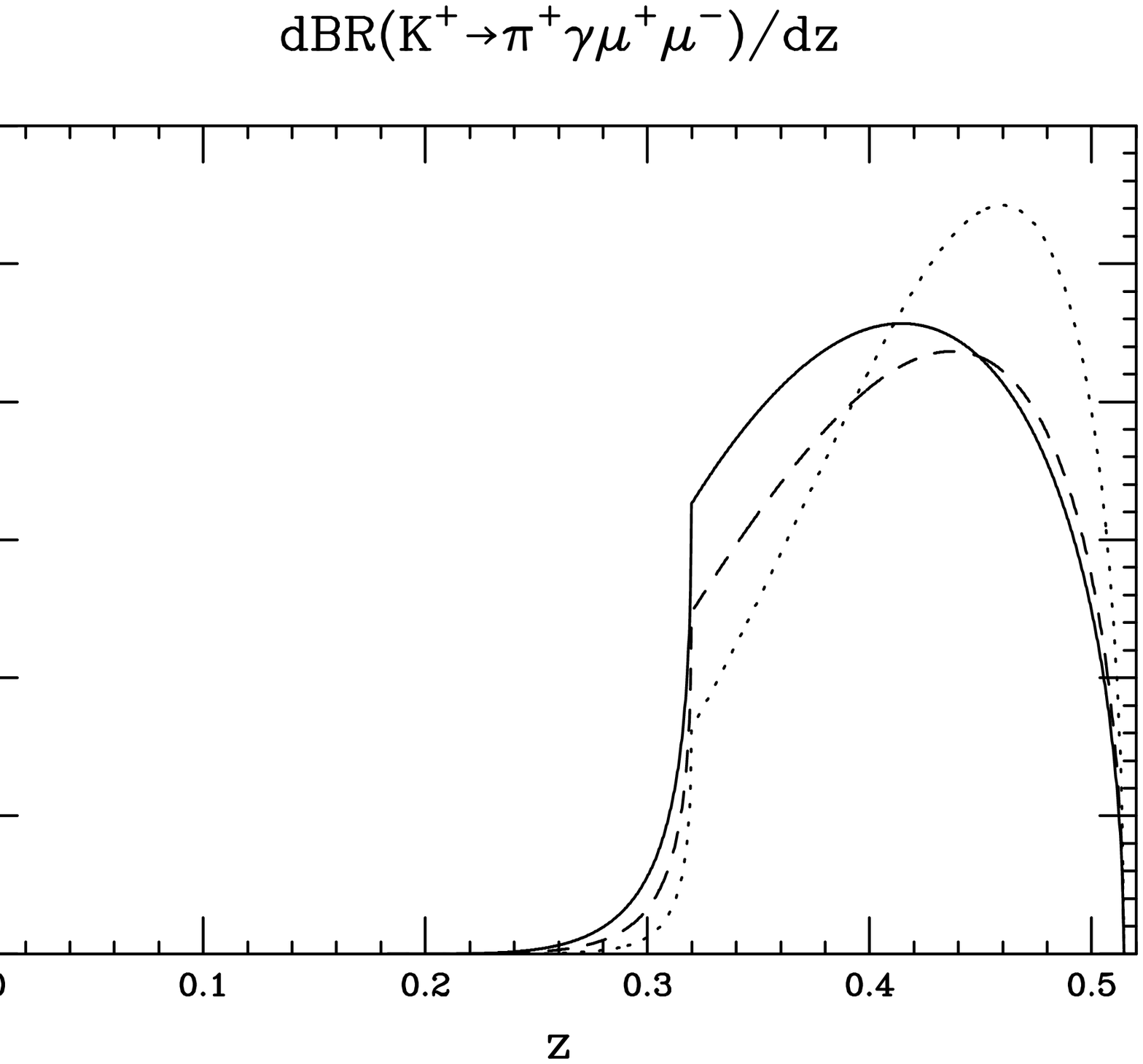,width=7.0cm}}
\caption{The differential branching ratio dBR(\kpgmm)/dz
to order E$^4$ is plotted
vs $z$ for $\hat c$ = 1.8 (solid line), $\hat c$ =
0 (dashed line) and $\hat c$ = --2.3 (dotted line).}
\end{minipage}
\hspace{1.0cm}
\begin{minipage}[t]{.44\linewidth}\centering
\mbox{\epsfig{file=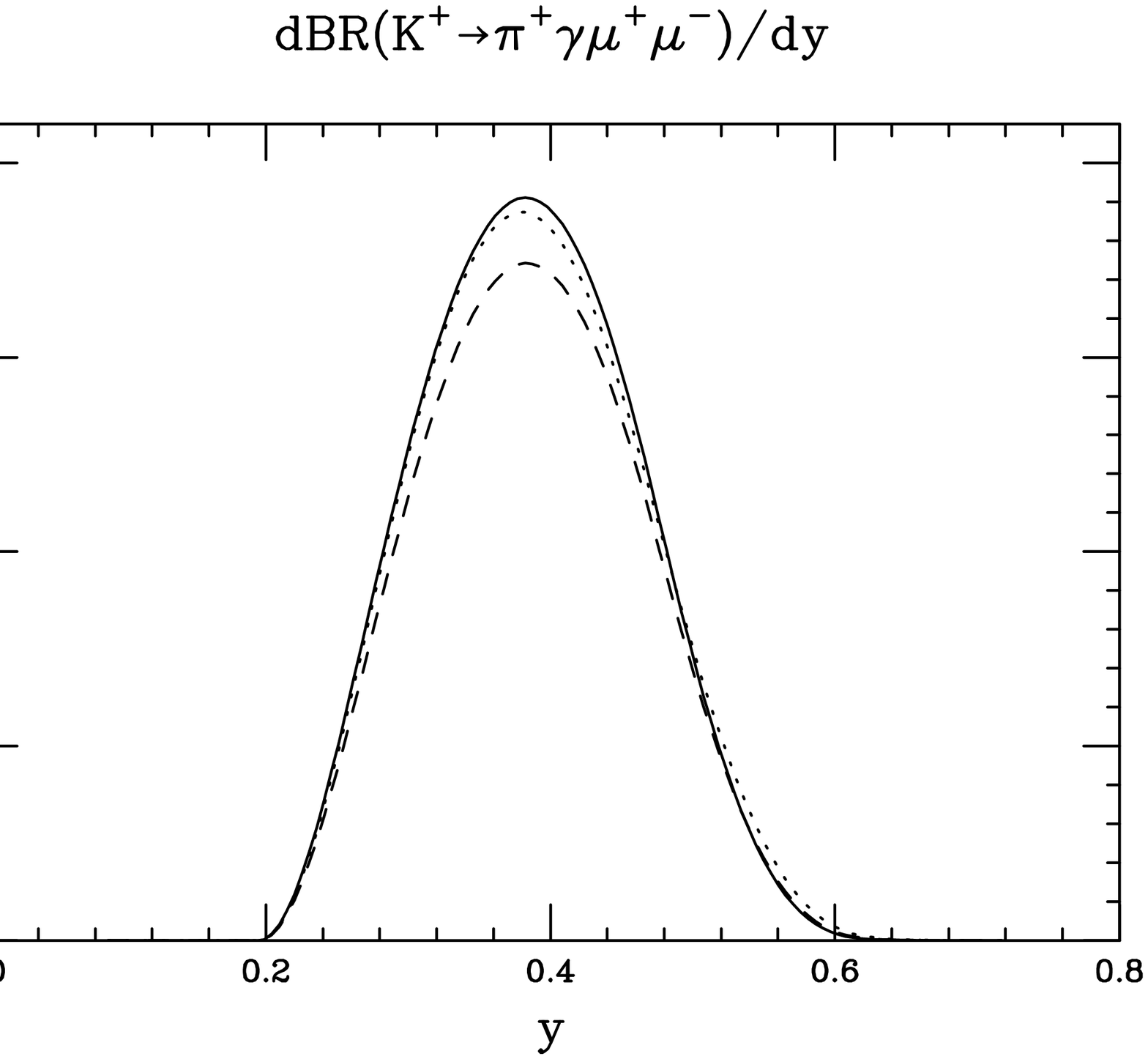,width=7.0cm}}
\caption{The differential branching ratio
dBR(\kpgmm)/dy
to order E$^4$ is plotted vs $y$ for $\hat c$ = 1.8 (solid line), $\hat c$ =
0 (dashed line) and $\hat c$ = --2.3 (dotted line).}
\end{minipage}}
\end{figure}

\section{\mbox{\boldmath $\cal{O}$(E$^6$) calculation}}

In this section we extend this calculation along the lines proposed by
the authors of Refs. \cite{CEP,CDM} for $K_L$ decays and
D'Ambrosio and Portol\'es for $K^+$ decays \cite{DP}. The former
provided a plausible solution to the problem raised by the
experimental rate not agreeing with the $\cal{O}$(E$^4$) calculation
when both photons are on shell. We have to add a new ingredient
that involves known physics that surfaces at the next order in the energy
expansion, i.e. the known quadratic energy variation of
the $K \ri 3\pi$ amplitude, which occurs from higher order terms in
the weak nonleptonic Lagrangian \cite{DGH,CDM,DEIN,DI}. While the
full one-loop structure of this is known \cite{KMW1,KMW2,DH}, it
involves complicated nonanalytic functions and we approximate the
result at $\cal{O}$(E$^4$) by an analytic polynomial which provides a
good description of the data throughout the physical region
\cite{ZE,KMW2}. Expanding in powers of the Dalitz plot variables,

\begin{eqnarray}
A^{(4)}(K^+ \ri \pi^+ \pi^+ \pi^-) & = &
2 \alpha_1 - \alpha_3 + \left( \beta_1 -
\Frac{1}{2} \beta_3 + \sqrt{3} \gamma_3 \right) Y \no \\
&-& 2 ( \zeta_1 + \zeta_3 ) \left( Y^2
+ \Frac{X^2}{3}\right) - ( \xi_1 + \xi_3 -
\xi_3')\left(Y^2 - \Frac{X^2}{3} \right). \label{fit}
\end{eqnarray}

\n Here the subscripts $1$ and $3$ refer to $\Delta I = 1/2 , 3/2$
transitions respectively, and the coefficients in Eq. (\ref{fit}) have
been fitted to the data \cite{KMW2}. We omit the $\Delta I = 3/2$
couplings $\zeta_3$ and $\xi_3$, $\xi_3'$ because of their big errors
shown in the fits in Ref. \cite{KMW2}. The Dalitz plot variables are
commonly defined as

\begin{equation}
X  =  \Frac{s_2 - s_1}{m_{\pi}^2}, \qquad
Y  =  \Frac{s_3 - s_0}{m_{\pi}^2},
\end{equation}

\n with $s_i = (p_K - p_i)^2$ for $i=1,2,3$, $s_0 =
(s_1 + s_2 + s_3)/3$, and the subscript $3$ indicates the odd pion
($\pi^0$ for $K_L$ decays and $\pi^-$ for $K^+$ decays).

In principle one can add the ingredients to the amplitudes and perform
a dispersive calculation of the total transition matrix element. In
practice it is simpler to convert the problem into an effective field
theory and and do a Feynman-diagram calculation which will yield the
same result. We follow this latter procedure.

The Feynman diagrams are the same as shown in Fig. 1, although the
vertices are modified by the presence of $\cal{O}$(E$^4$) terms in the
energy expansion. Not only does the direct $K \ri 3\pi$ vertex change
to the form given in Eq. (\ref{fit}), but also the weak vertices with
one and two photons have a related change. The easiest way to
determine these is to write a gauge invariant effective Lagrangian
with coefficients adjusted to reproduce Eq. (\ref{fit}). One also
has to add diagrams with one or two photons radiating from the
incoming $K^+$ or the outcoming $\pi^+$, or with one photon radiating
from one of the outcoming leptons.
This causes infrared divergences, which are to be treated in the
usual way, as part of a general calculation of radiative corrections
to the process $K^+ \ri \pi^+ l^+ l^-$. In practice, with an
appropriate set of experimental cuts on the phase space parameters, it
is possible to restrict the outcome to a measurable non-bremsstrahlung
contribution only \cite{AP}. Below we shall give an example of these
cuts and a prediction for the experimental result once they are implemented.

The resulting calculation follows the same steps as described in
Sec. III, but is more involved and is not easy to present in a
simple form. We have checked that our result reduces to that of
Ref. \cite{DP} in the limit of on-shell photons. Remembering the
definitions

\begin{equation}
r_{\pi} = {m_{\pi} \over m_K}, \qquad r_{\eta} = {m_{\eta} \over m_K},
\qquad z = {s \over {m^2_K}},
\qquad q = {{k^2_1} \over {m^2_K}},
\end{equation}

\n the unitarity one-loop corrections yield the following:

\begin{eqnarray}
{\cal M}_{\mu\nu} &=& {\alpha_{em} \over {2 \pi}}
\left[A(z,y,q)(k_{2\mu} k_{1\nu}-k_1 \cdot k_2 g_{\mu\nu})
\phantom{{k^2_1} \over {k_1}} \right. \no \\
\hskip -4pt &+& \hskip -4pt B(z,y,q) \left({p^{\phantom{l}}_K \cdot k_1
p^{\phantom{l}}_K \cdot k_2 \over {k_1 \cdot k_2}}g_{\mu\nu}+
p^{\phantom{l}}_{K\mu}p_{K\nu}-{{p^{\phantom{l}}_K \cdot k_1}
\over {k_1 \cdot k_2}} k_{2\mu} p^{\phantom{l}}_{K\nu}-
{{p^{\phantom{l}}_K \cdot k_2} \over {k_1 \cdot k_2}}
k_{1\nu} p^{\phantom{l}}_{K\mu}\right) \no \\
&+& C_1(z) \varepsilon_{\mu \nu \rho \sigma}
{k_1}^{\rho} {k_2}^{\sigma} \no \\
\hskip -4pt &+& \hskip -4pt \left. D(z,y,q) \left(k^2_1
{p^{\phantom{l}}_K \cdot k_2
\over {k_1 \cdot k_2}}g_{\mu\nu}-
{{p^{\phantom{l}}_K \cdot k_2} \over {k_1 \cdot k_2}} k_{1\mu} k_{1\nu}+
k_{1\mu}p^{\phantom{l}}_{K\nu}-{{k^2_1} \over {k_1 \cdot k_2}}k_{2\mu}
p^{\phantom{l}}_{K\nu}\right) \right], \no \\
\phantom{l}
\end{eqnarray}

\n where

\begin{eqnarray}
A m^2_K &=& {{G_8 m^2_K}\over{(z-q)}} \{(z+r^2_{\pi}-1)[1+2 I(m^2_K)]
-\hat{c}(z-q)\}\no \\&+&
\left\{2(2 \alpha_1 - \alpha_3)+\left(1 + {1 \over {3 r^2_{\pi}}}-{z \over
r^2_{\pi}}\right)\left(\beta_1 - {1 \over 2} \beta_3 +
\sqrt{3}\gamma_3\right)\no \right. \\&-&
\left. {8 \over {3 r^4_{\pi}}}{(2\zeta_1-\xi_1)}{1 \over
{18}}\left[1+6(r^2_{\pi}-z)+9(r^2_{\pi}-z)^2\right]\right\}[1+2 I(m^2_{\pi})]
\no \\ &-&
{8 \over {3 r^4_{\pi}}}{(2\zeta_1-\xi_1)}
\left[\left(r^2_{\pi}-{q \over 12}\right)
\log{m^2_{\pi} \over {\mu^2}}+{1 \over 2}I_4\right] \no \\ &-&
\a2 \left\{ \phantom{m^2 \over \mu^2}\hskip -19pt - \{2
[1-2(x_1 + x_2)] I_1(z_1 z_2)+x_1 I_1(z_2)\right. \phantom{1 \over 2}\no \\
&+& x_2 [2 I_1(z^2_2) - I_1(z_2)+
I_1(z_1)]\} \no \\
&+& 2 \{[2 x^2_1 -x_1(z+q)]
[-I_2(z^3_1 z_2)+I_2(z^2_1 z_2)] \no \\
&+& [2 x_1 x_2 - x_1 (z-q)/2 -x_2 (z+q)/2]
[2 I_2(z^2_1 z^2_2)+I_2(z_1 z_2)-I_2(z^2_1 z_2) \no \\
&-& I_2(z_1 z^2_2)]+[2 x^2_2-x_2 (z-q)]
[I_2(z_1 z^2_2)-I_2(z_1 z^3_2)]\} \no \\
&+& \left[{1 \over 9}(1 - 3 r^2_{\pi})+{1 \over 12}r^2_{\pi}(1+3 r^2_{\pi})
\left(1 + {1 \over {3 r^2_{\pi}}}-{z \over r^2_{\pi}}\right)
\right][1+2I(m^2_{\pi})] \no \\
&-& \left. {1 \over 12} \left(1 + \log{m^2_{\pi} \over
{\mu^2}}\right)-{1 \over 2}\left(r^2_{\pi}-{q \over 12}\right)
\log{m^2_{\pi} \over {\mu^2}}-{1 \over 4}I_4\right\},
\end{eqnarray}

\begin{eqnarray}
B m^2_K &=& \a2 \left\{-2 I_3 +I_4 +
{1 \over 12} (z-q) \log{m^2_{\pi} \over {\mu^2}}
\right. \no \\
&-& \left. {1 \over 4} \left({q \over 6}\log{m^2_{\pi} \over
{\mu^2}} - I_4\right) \right\},
\end{eqnarray}

\begin{equation}
C_1 m^2_K = {2 G_8 m^2_K}{{2+r^2_{\pi}-3
r^2_{\eta}} \over {3 (z-r^2_{\pi})}},
\end{equation}

\begin{eqnarray}
D m^2_K &=& \a2 \left\{I_3 -{I_4 \over 2}
- {1 \over 24} (z-q) \log{m^2_{\pi} \over {\mu^2}} \right.
\no \\
&+& [2 x_2 -(z-q)/2][2 I_1(z_1 z_2)-I_1(z_2)] \no \\
&+& \left. (2 y -q)[I_1(z_1)-I_1(1)/2]+[2 x_1-(z+q)/2]
{I_5 \over 4} \right\}.
\end{eqnarray}

\n The integrals used in the above formulas are defined here
and given explicitly in the Appendix:

\begin{equation}
I_1(z^n_1 z^m_2) = \int^1_0 dz_1 \int^{1-z_1}_0 dz_2 z^n_1 z^m_2
\log{D_1 \over {m^2_{\pi}}},
\end{equation}

\begin{equation}
{I_2(z^n_1 z^m_2) \over m^2_K} = \int^1_0 dz_1
\int^{1-z_1}_0 dz_2 {{z^n_1 z^m_2} \over D_1},
\end{equation}

\begin{equation}
I_3 m^2_K= \int^1_0 dz_1 \int^{1-z_1}_0 dz_2 D_1 \log{D_1 \over {m^2_{\pi}}},
\end{equation}

\begin{equation}
I_4 m^2_K = \int^1_0 dz_1 D_2 \log{D_2 \over {m^2_{\pi}}},
\end{equation}

\begin{equation}
I_5 = \int^1_0 dz_1 (4 z^2_1-4 z_1+1) \log{D_2 \over {m^2_{\pi}}},
\end{equation}

\n where

\begin{eqnarray}
D_1 &=& m^2_{\pi} - 2 k_1 \cdot k_2 z_1 z_2 - k^2_1 z_1 (1-z_1),
\no \\
D_2 &=& m^2_{\pi} - k^2_1 z_1 (1-z_1), \no \\
x_1 &=& {{p^{\phantom{l}}_K \cdot k_1} \over {m^2_K}},
\qquad x_2 = {{p^{\phantom{l}}_K \cdot k_2} \over {m^2_K}}.
\end{eqnarray}

The above formulas lead to the total branching ratios shown in Table II,
in full analogy with the results of Sec. III. The numerical results
are obtained for the mass scale $\mu$ = $m_{\rho}$ and setting all the
counterterms to 0 \cite{DP}. The corresponding decay distributions are
plotted in Figs. 6--9.

\begin{table}[htbp]
\centering
\caption{Results for BR($K^+\ri\pi^+\gamma l^+l^-$) at ${\cal O}$(E$^6)$.}
\vskip 0.1 in
\begin{tabular}{ccc} \hline\hline\\
& BR(\kpgee) & BR(\kpgmm) \\
\hline\\
$\hat{c}$ = 1.8 (fit) & 1.7 $\times$ 10$^{-8}$ & 7.0 $\times$ 10$^{-11}$ \\
$\hat{c}$ = 0 (WDM) & 1.1 $\times$ 10$^{-8}$ & 7.3 $\times$ 10$^{-11}$ \\
$\hat{c}$ = --2.3 (FM) & 9.2 $\times$ 10$^{-9}$ & 8.5 $\times$ 10$^{-11}$ \\
\hline\hline
\end{tabular}
\end{table}

\begin{figure}[htbp]
\vfill
\centerline{
\begin{minipage}[t]{.44\linewidth}\centering
\mbox{\epsfig{file=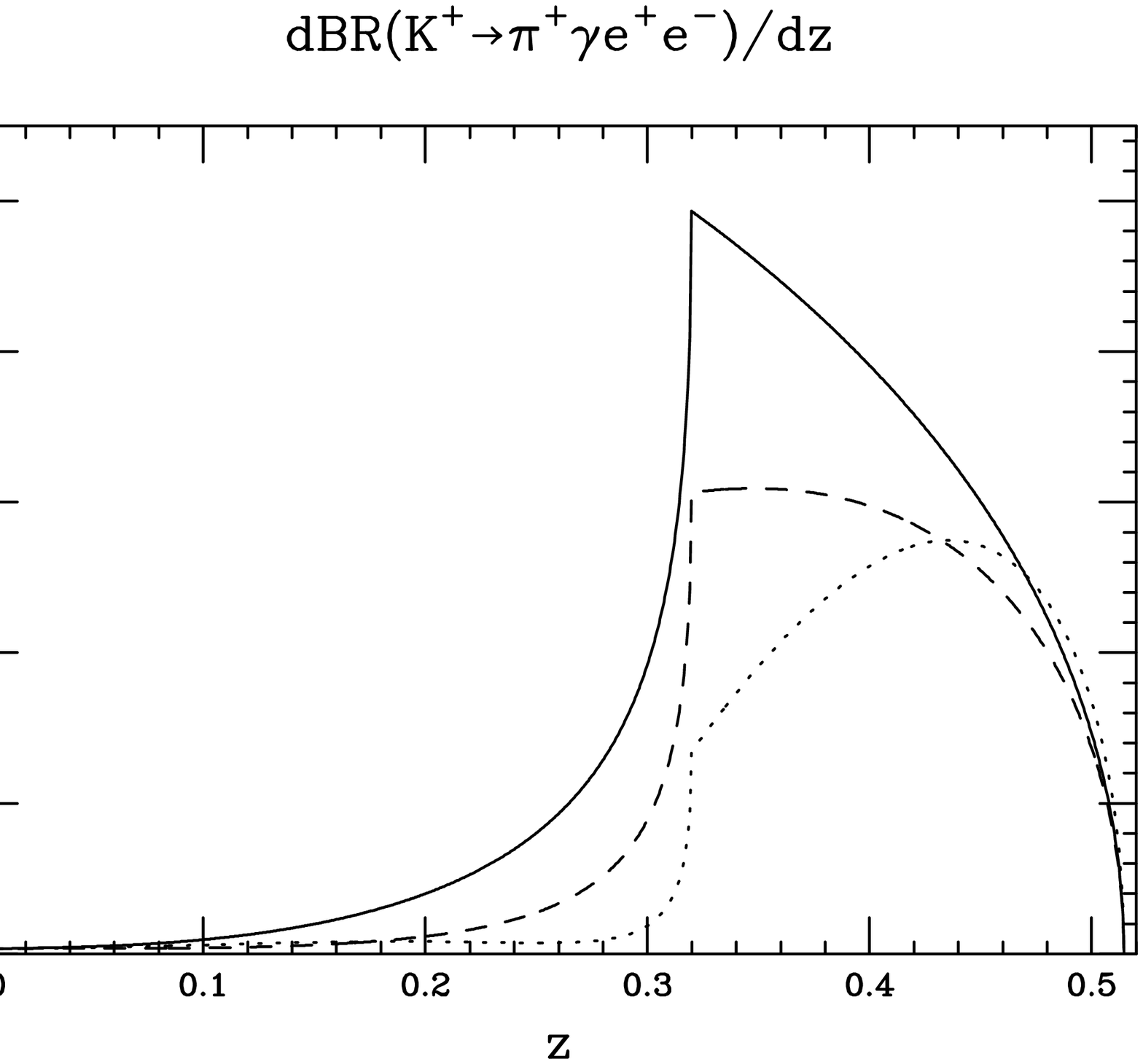,width=7.0cm}}
\caption{The differential branching ratio
dBR(\kpgee)/dz to order E$^6$ is plotted
vs $z$ for $\hat c$ = 1.8 (solid line), $\hat c$ = 0 (dashed line)
and $\hat c$ = --2.3 (dotted line).}
\end{minipage}
\hspace{1.0cm}
\begin{minipage}[t]{.44\linewidth}\centering
\mbox{\epsfig{file=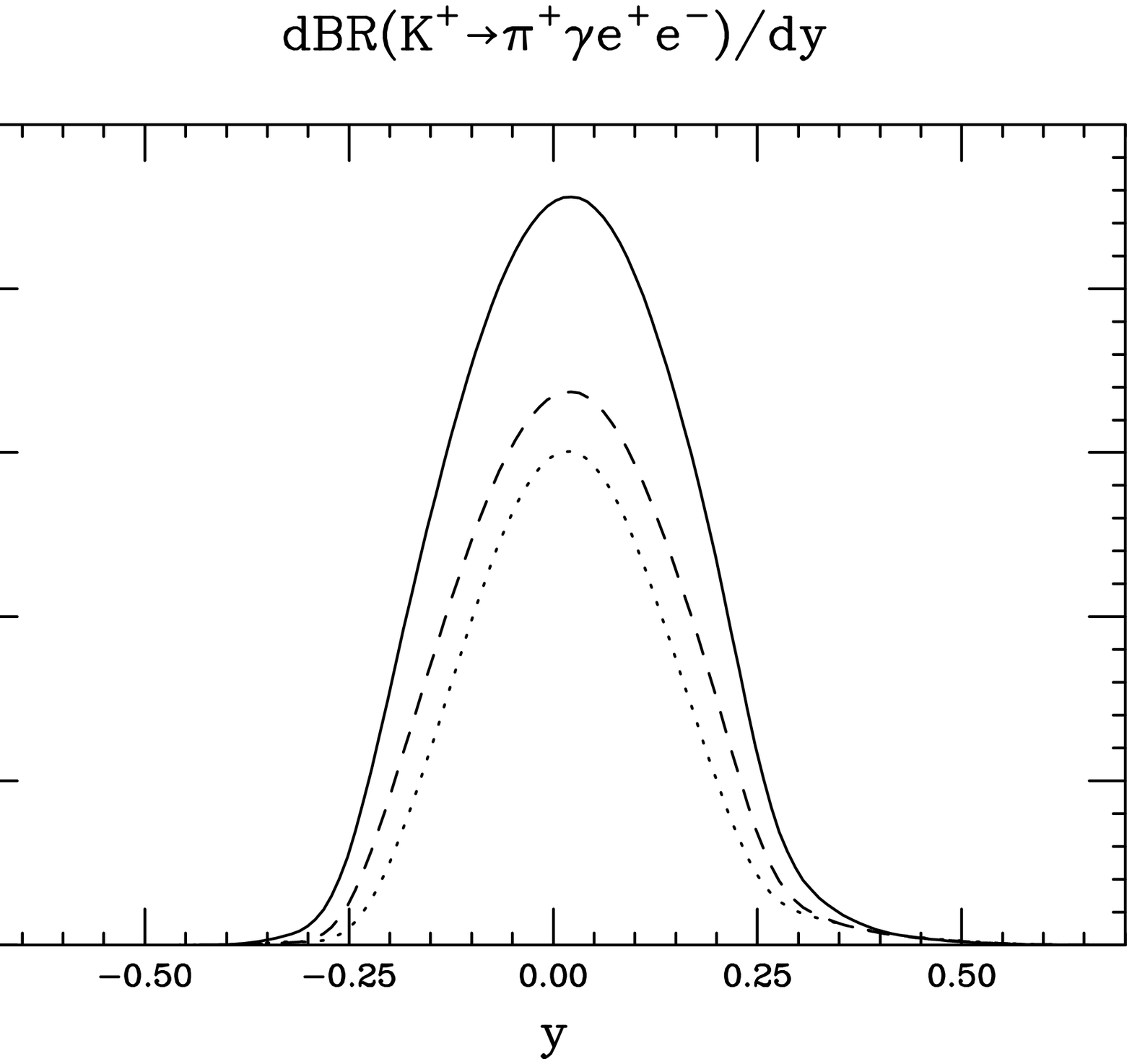,width=7.0cm}}
\caption{The differential branching ratio
dBR(\kpgee)/dy to order E$^6$ is plotted
vs $y$ for $\hat c$ = 1.8 (solid line), $\hat c$ = 0 (dashed line)
and $\hat c$ = --2.3 (dotted line).}
\end{minipage}}
\end{figure}

\begin{figure}[htbp]
\vfill
\centerline{
\begin{minipage}[t]{.44\linewidth}\centering
\mbox{\epsfig{file=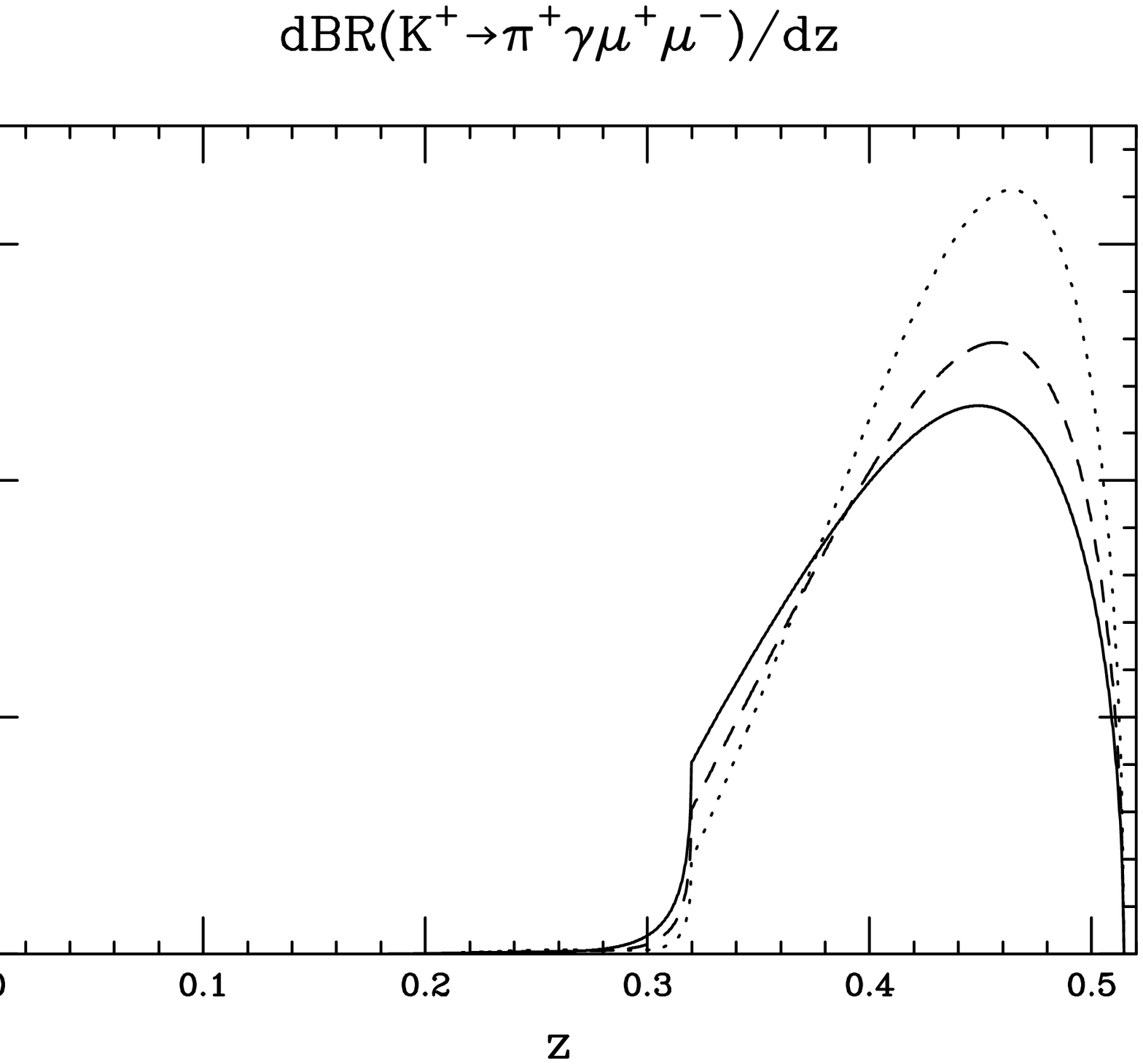,width=7.0cm}}
\caption{The differential branching ratio
dBR(\kpgmm)/dz to order E$^6$ is plotted
vs $z$ for $\hat c$ = 1.8 (solid line), $\hat c$ = 0 (dashed line)
and $\hat c$ = --2.3 (dotted line).}
\end{minipage}
\hspace{1.0cm}
\begin{minipage}[t]{.44\linewidth}\centering
\mbox{\epsfig{file=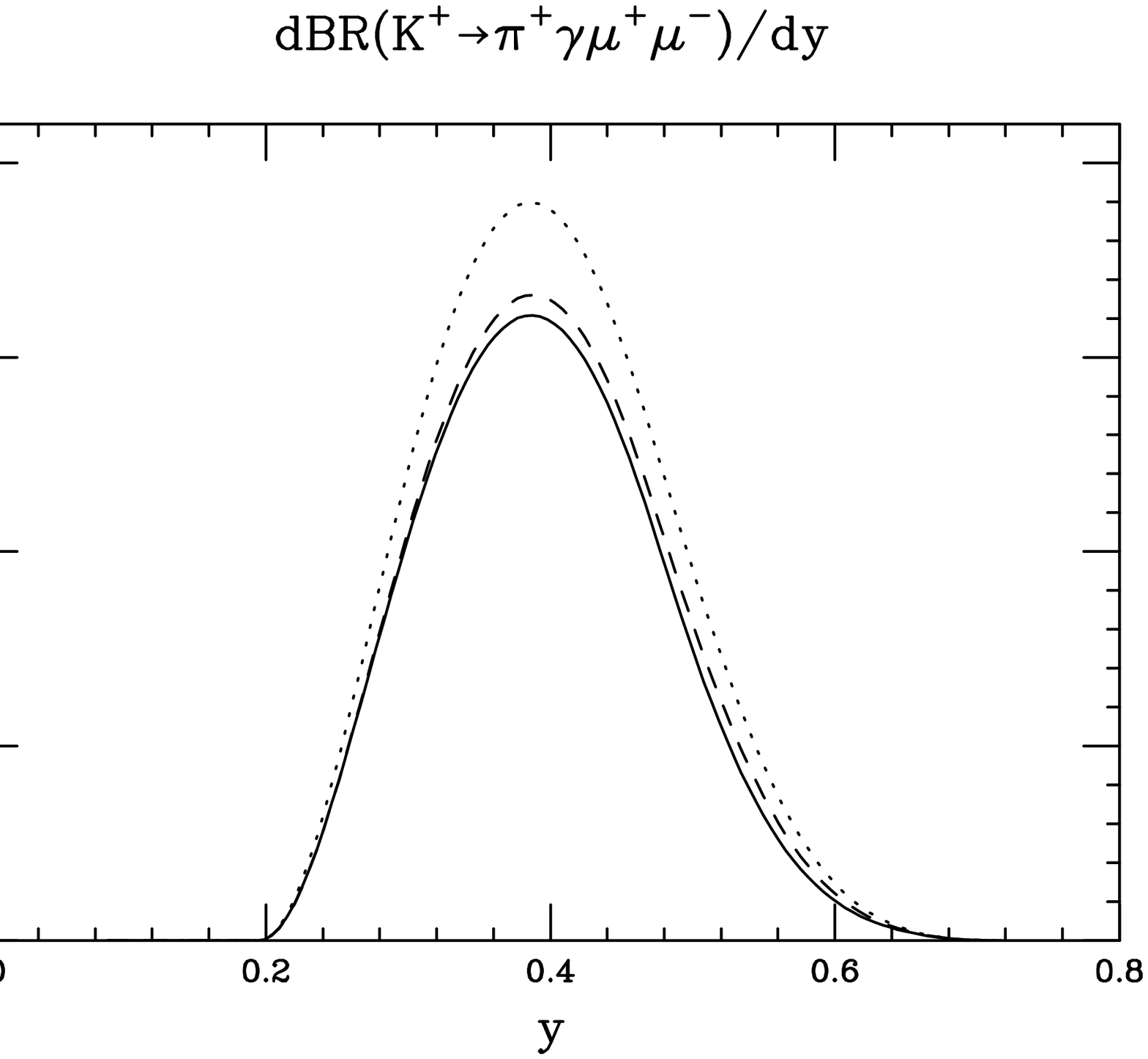,width=7.0cm}}
\caption{The differential branching ratio
dBR(\kpgmm)/dy to order E$^6$ is plotted
vs $y$ for $\hat c$ = 1.8 (solid line), $\hat c$ = 0 (dashed line)
and $\hat c$ = --2.3 (dotted line).}
\end{minipage}}
\end{figure}

The uncertainty in the theoretical prediction is dominated by the
unknown ${\cal O}($E$^4)$ counterterm generated amplitude $\hat{c}$ in
Eq. (\ref{a4}). In Figs. 10 and 11 we plot BR(\kpgee) and BR(\kpgmm) as
a function of $\hat{c}$, both with and without the ${\cal O}($E$^6)$
corrections just computed.

\begin{figure}[htbp]
\vfill
\centerline{
\begin{minipage}[t]{.44\linewidth}\centering
\mbox{\epsfig{file=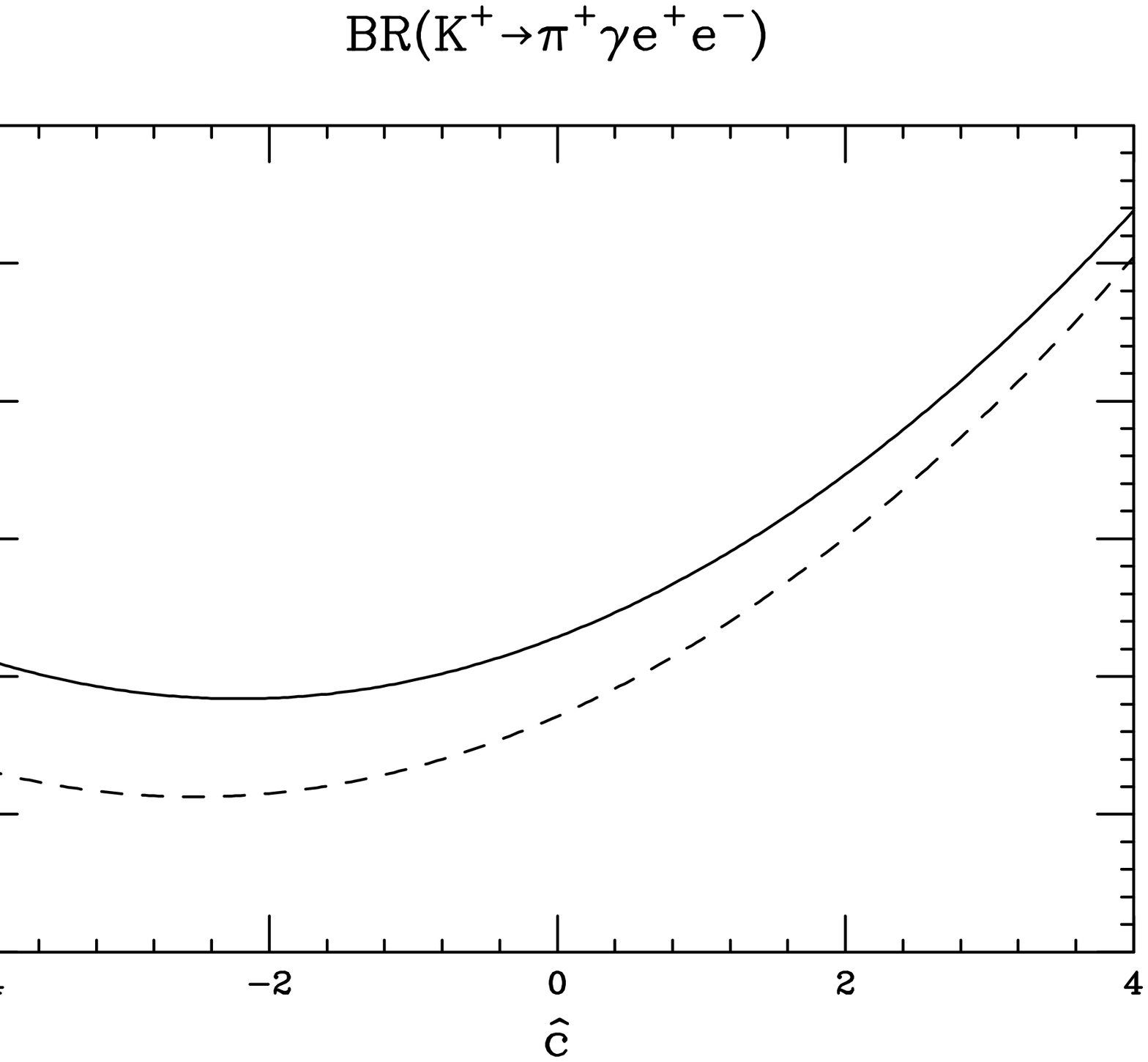,width=7.0cm}}
\caption{The branching ratio BR(\kpgee) is
plotted vs $\hat c$ at ${\cal O}($E$^4$) (dashed line) and up to
${\cal O}($E$^6$) (solid line).}
\end{minipage}
\hspace{1.0cm}
\begin{minipage}[t]{.44\linewidth}\centering
\mbox{\epsfig{file=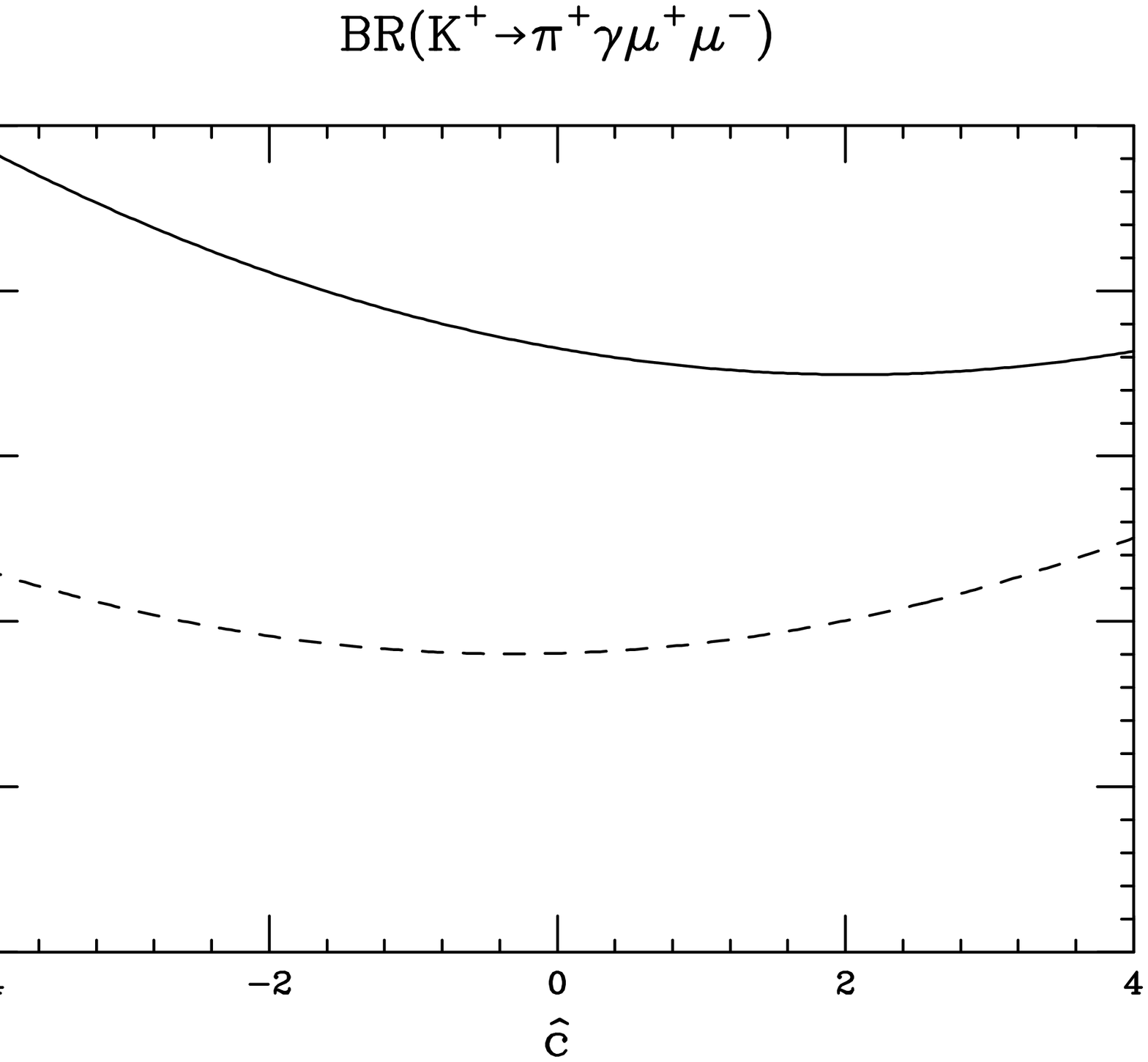,width=7.0cm}}
\caption{The branching ratio BR(\kpgmm) is
plotted vs $\hat c$ at ${\cal O}($E$^4$) (dashed line) and up to
${\cal O}($E$^6$) (solid line).}
\end{minipage}}
\end{figure}

If we implement the experimental cuts currently used at BNL to extract
the non-bremsstrahlung contribution to \kpgee\ \cite{AP},

\begin{equation}
m_{e^+e^-} \geq 150\ {\rm MeV}, 
\quad E_{\gamma}\geq 30\ {\rm MeV},
\quad m_{e^+\gamma}, m_{e^-\gamma} \geq 30\  {\rm MeV},
\label{cuts}
\end{equation}

\n the resulting theoretical branching ratios are reduced by more than
an order of magnitude. They are presented in Table III. Preliminary
experimental data are not conclusive at the present stage. In Figs. 12
and 13 we plot the differential branching ratios up to ${\cal O}$(E$^6)$,
taking into account the above cuts in the phase space integration.

\begin{table}[htbp]
\centering
\caption{Results for BR($K^+\ri\pi^+\gamma e^+ e^-$) at ${\cal
O}$(E$^6)$ with the cuts defined by (\ref{cuts}).}
\vskip 0.1 in
\begin{tabular}{ccc} \hline\hline\\ 
& BR(\kpgee)\\
\hline\\
$\hat{c}$ = 1.8 (fit) & 5.5 $\times$ 10$^{-10}$ \\
$\hat{c}$ = 0 (WDM) & 4.8 $\times$ 10$^{-10}$ \\
$\hat{c}$ = --2.3 (FM) & 4.7 $\times$ 10$^{-10}$ \\
\hline\hline
\end{tabular}
\end{table}

\begin{figure}[htbp]
\vfill
\centerline{
\begin{minipage}[t]{.44\linewidth}\centering
\mbox{\epsfig{file=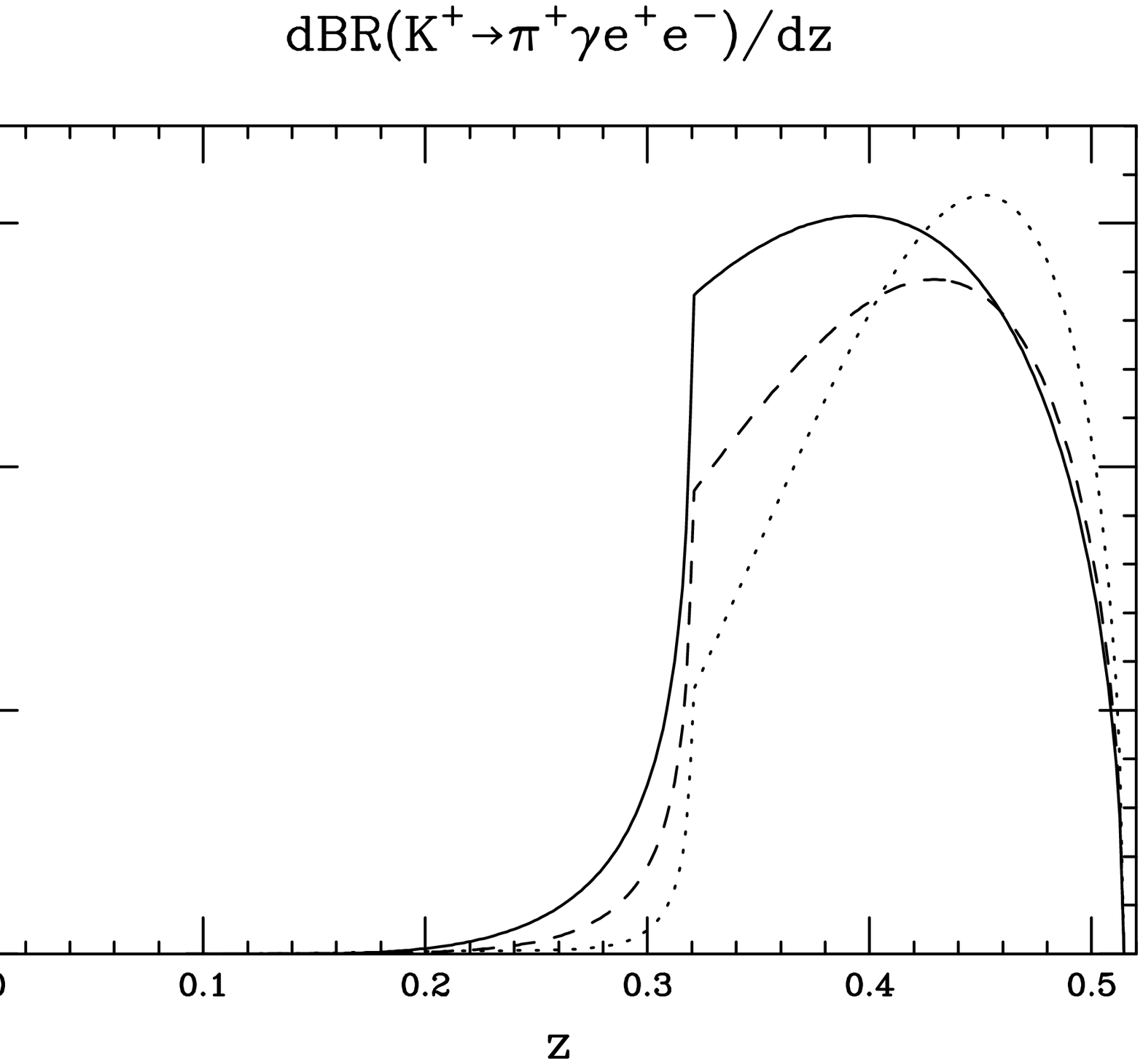,width=7.0cm}}
\caption{The differential branching ratio
dBR(\kpgee)/dz to order E$^6$ is plotted
vs $z$ for $\hat c$ = 1.8 (solid line), $\hat c$ = 0 (dashed line)
and $\hat c$ = --2.3 (dotted line) with the cuts defined by the
inequalities (\ref{cuts}).}
\end{minipage}
\hspace{1.0cm}
\begin{minipage}[t]{.44\linewidth}\centering
\mbox{\epsfig{file=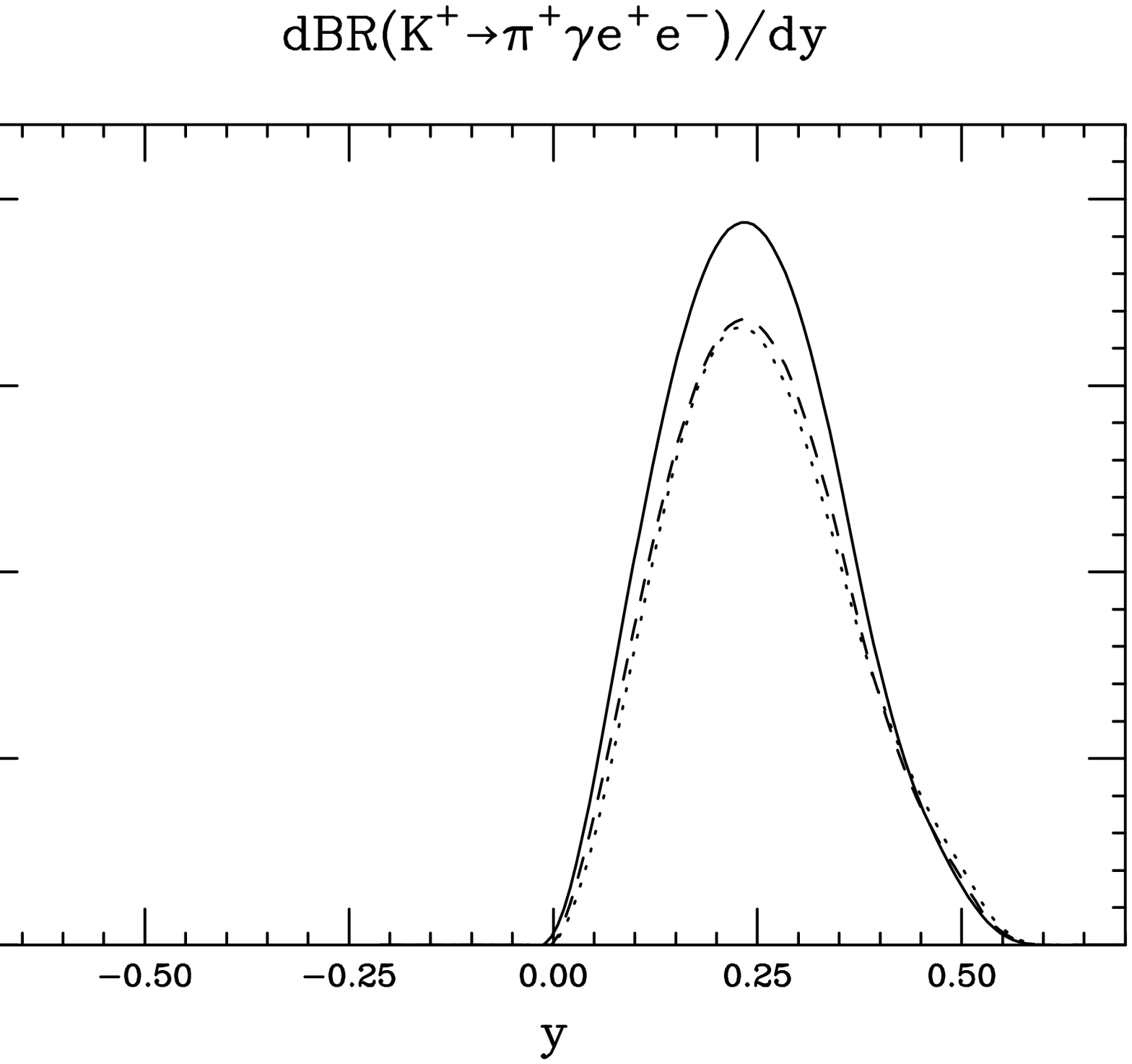,width=7.0cm}}
\caption{Same as in Fig. 12 for the differential branching ratio
dBR(\kpgee)/dy vs $y$.}
\end{minipage}}
\end{figure}

\section{Conclusions}

We have computed the unitarity corrections at one loop
in ChPT for the processes \kpgee\ and \kpgmm, allowing us to present
predictions for their total and differential branching ratios.

As expected, the muonic rate is significantly smaller than in the
corresponding electronic mode, analogously to the cases studied in
Refs. \cite{DG1} and \cite{DG2}. This is of course due to the more
limited phase space, as well as the fact that the photon propagator is
further off shell in the muonic case. We again see that the more
complete calculation presented above leads to a conspicuous
enhancement over the purely order E$^4$ calculation presented
first. The vector meson diagrams here are not expected to add a
significant amount to the overall rates and have been omitted
altogether. Their inclusion provided a consistent contribution to
the $K_L$ decays, leading to dramatic increases in the ${\cal O}$(E$^6$)
results. Therefore, in the computation presented in this paper the
enhancement is expected to be somewhat smaller.

The results for the differential branching ratios follow a
pattern recognizable in all the previous calculations for radiative
rare kaon decays: A large peak is visible above the two-pion
threshold in the $z$ variable, with a tail extended to low $z$, and a
slightly asymmetrical and structureless distribution in the $y$ variable.
Experimentally, the abundance of information supplied by the former
plots makes them a preferred option (see for example
\cite{KTEV1}). This is true also if we reduce the phase space of
integration performing experimental cuts.

We also note that comparing Figs. 10 and 11, one sees that the
dependence on $\hat{c}$ is markedly higher when electrons, instead of
muons, are present in the final state. This has important consequences
if one tries to extract the value of $\hat{c}$ from the data.

\bigskip\bigskip
\n {\Large \bf Acknowledgments}
\bigskip

\n This work is supported in part by the U.S. Department of Energy
under Grant DE-FG05-96ER40945.

\appendix
\newcounter{homero}
\renewcommand{\thesection}{\Alph{homero}}
\renewcommand{\theequation}{\Alph{homero}\arabic{equation}}
\setcounter{homero}{1}
\setcounter{equation}{0}

\section*{Appendix: Relevant integrals}

In this appendix we list the explicit expressions for the integrals
used in the calculation of Sec. IV. We follow the notation of that
section. For $s \leq 4 \mp^2$ and $\k1^2 \leq 4 \mp^2$ we have

\begin{eqnarray}
\label{first}
I_1(1) &=& {1 \over {s-k^2_1}}\left\{ -{3 \over 2}\left(s-\k1^2\right)-\mp^2
\left[F(s) -F(\k1^2)\right] \phantom{\ak} \right. \no \\
&-& \left. \ka \ak + \sa \as \right\}, \no \\
\end{eqnarray}

\begin{eqnarray}
I_1(z_1) &=& {1 \over {s-k^2_1}}\left[-{{4 s} \over 9} - {(4 m^2_{\pi}- s)
\over {3 \sqrt{s}}} \sa \as\right. \no \\
&+& \left. {{4 k^2_1} \over 9} + {(4 m^2_{\pi}- k^2_1)
\over {3 \sqrt{k^2_1}}} \ka \ak \right],
\end{eqnarray}

\begin{eqnarray}
I_1(z^2_1) &=& {1 \over {s-k^2_1}}\left[-{{2 s} \over 9} - {(4 m^2_{\pi}- s)
\over {6 \sqrt{s}}} \sa \as \right. \no \\
&+& \left. {{2 k^2_1} \over 9} + {(4 m^2_{\pi}- k^2_1)
\over {6 \sqrt{k^2_1}}} \ka \ak \right],
\end{eqnarray}

\begin{eqnarray}
I_1(z_2) &=& -{4 \over 9} + {{3 m^2_{\pi} + k^2_1} \over {6 \ds}} +
{m^2_{\pi} \over {3 s}} + {{(-4 m^4_{\pi} + 5 m^2_{\pi} s -s^2)} \over
{3 s \sqrt{s} \sa }} \as \no \\
&-& {m^4_{\pi} \over {2 \ds^2}} \left[ 2 {\sa \over m^2_{\pi}} \as -
{s \over m^2_{\pi}} \right. \no \\
&+& \left. 2 {\ka \over m^2_{\pi}} \ak +
{k \over m^2_{\pi}}\right] \no \\
&-& {k^4_1 \over {3 \ds^2}} \left[{m^2_{\pi} \over s} + {{(-4 m^4_{\pi}
+ 5 m^2_{\pi} s -s^2)} \over {s \sqrt{s} \sa}} \as - {{m^2_{\pi}} \over
k^2_1} \right. \no \\
&+& \left. {{(-4 m^4_{\pi} + 5 m^2_{\pi} k^2_2 -k^4_1)} \over {k^2_1
\sqrt{k^2_1} \ka}} \ak \right] \no \\
&+& {m^2_{\pi} \k1^2 \over \ds^2}\left[F(s) - F(k^2_1)\right] - {{2 m^2_{\pi}
k^2_1} \over \ds^2}\left[{\sa \over \sqrt{s}} \as \right. \no \\
&-& \left. {\ka \over \sqrt{k^2_1}} \ak\right],
\end{eqnarray}

\begin{eqnarray}
I_1(z^2_2) &=& -{2 \over 9} + {m^2_{\pi} \over {4 \ds}} + {k^2_1 \over
{24 \ds}} + {1 \over {3 \ds^2}}\left(m^4_{\pi} - 3 m^2_{\pi} k^2_1 -
{\k1^4 \over 4}\right) \no \\
&+& {\mp^2 \over {3s}}-{\nns \over {6 s \sqrt{s} \sa}} \as \no \\
&-& {\mp^6 \over {3 \ds^3}} \left[ {s \over \mp^2} + {{s \sqrt{s} \sa}
\over \mp^4} \as \right. \no \\
&-& \left. {\k1^2 \over \mp^2} - {{\k1^2 \sqrt{\k1^2} \ka}
\over \mp^4} \ak \right] \no \\
&+& {\k1^6 \over {3 \ds^3}} \left[{\mp^2 \over s} -{\nns \over {2s
\sqrt{s}\sa}} \as \right. \no \\
&-& \left. {\mp^2 \over \k1^2} +{\nnk \over {2 \k1^2
\sqrt{\k1^2}\ka}} \ak \right] \no \\
&+& {{\mp^4 \k1^2} \over {\ds^3}} \left[ -{s \over \mp^2} + {{2 \sa \sqrt{s}}
\over \mp^2} \as \right. \no \\
&+& \left. {\k1^2 \over \mp^2} - {{2 \ka \sqrt{\k1^2}}
\over \mp^2} \ak - F(s) + F(\k1^2) \right] \no \\
&-& {{\mp^2 \k1^4} \over {\ds^3}} \left[ F(s) -{{3 \sa}
\over \sqrt{s}} \as \right. \no \\
&-& \left. F(\k1^2) + {{3 \ka}
\over \sqrt{\k1^2}} \ak \right],
\end{eqnarray}

\begin{eqnarray}
I_1(z_1 z_2) &=& -{13 \over 144}+{{(-6 m^2_{\pi}+k^2_1)} \over {24 \ds}}
+ {\ns \over {12 s \sqrt{s} \sa}} \as \no \\
&+& {m^4_{\pi} \over {2 \ds^2}}\left[F(k^2_1)-F(s)\right] \no \\
&-& {k^4_1 \over {2 \ds^2}} \left[- {{m^2_{\pi}} \over {3 s}} +
{\ns \over {6 s \sqrt{s} \sa}}\as \right. \no \\
&+& \left. {{m^2_{\pi}} \over {3 k^2_1}} - {\nk \over {6 k^2_1
\sqrt{k^2_1} \ka}}\ak \right] \no \\
&+& {{m^2_{\pi} k^2_1} \over \ds^2} \left[ {\sa \over \sqrt{s}}\as
\right. \no \\
&-& \left. {\ka \over \sqrt{\k1^2}}\ak \right],
\end{eqnarray}

\begin{eqnarray}
{I_2(z_1 z_2) \over m^2_K} &=& -{1 \over {2 \ds}} -{\mp^2 \over \ds^2}
\left[ F(s) - F(\k1^2)\right] \no \\ &+& {\k1^2 \over \ds^2}
\left[ {\sa \over \sqrt{s}} \as
\right. \no \\
&-& \left. {\ka \over \sqrt{\k1^2}} \ak \right],
\end{eqnarray}

\begin{eqnarray}
{I_2(z_1 z^2_2) \over m^2_K} &=& -{1 \over {6 \ds}} + {1 \over
\ds^2}\left({\k1^2 \over 3} + \mp^2\right) \no \\ &-& {\mp^2
\over \ds^3}\left[2 \sa \sqrt{s} \as -s \right. \no \\ &-&
\left. 2 \ka \sqrt{\k1^2} \ak + \k1^2 \right] \no \\ &-& {{2
\k1^4} \over {3 \ds^3}}\left[{\mp^2 \over s}+ {{(-4 m^4_{\pi}+5
m^2_{\pi} s-s^2)} \over {s \sqrt{s} \sa}} \as
\right. \no \\
&-& \left. {\mp^2 \over \k1^2} - {{(-4 m^4_{\pi} + 5 m^2_{\pi} k^2_2 -k^4_1)}
\over {k^2_1 \sqrt{k^2_1} \ka}} \ak \right] \no \\
&+& {{2 \mp^2 \k1^2} \over \ds^3} \left[F(s) - F(\k1^2)
- {{2 \sa} \over \sqrt{s}}\as \right. \no \\
&+& \left. {{2 \ka} \over \sqrt{\k1^2}}\ak
\right],
\end{eqnarray}

\begin{eqnarray}
{I_2(z_1 z^3_2) \over m^2_K} &=& -{1 \over {12 \ds}} + {1 \over \ds^2}
\left({{6 \mp^2 +\k1^2} \over 4} \right)\no \\ &-& {1 \over
\ds^3} \left({{4 \mp^4 -12 \mp^2 \k1^2 -\k1^4} \over 4}\right)
\no \\ &-& {\mp^6 \over \ds^4}\left[{s \over \mp^2} - {\k1^2
\over \mp^2} + {{s \sqrt{s}\sa} \over \mp^4} \as \right. \no \\
&-& \left. {{\k1^2 \sqrt{\k1^2} \ka}
\over \mp^4} \ak - {s^2 \over {4 \mp^4}} + {\k1^4 \over
{4 \mp^4}} \right] \no \\
&+& {\k1^6 \over \ds^4}\left[{\mp^2 \over s} -{\nns \over {s
\sqrt{s}\sa}} \right. \as \no \\
&-& \left. {\mp^2 \over \k1^2} + {\nnk \over {\k1^2
\sqrt{\k1^2}\ka}} \ak \right] \no \\
&+& {{3 \mp^4 \k1^2} \over \ds^4} \left[ {{2 \sa \sqrt{s}}\over
{\mp^2}}\as - {s \over \mp^2}\right. \no \\
&+& \left. {\k1^2 \over \mp^2}- {{2 \ka \sqrt{\k1^2}}\over
{\mp^2}}\ak -F(s) + F(\k1^2) \right] \no \\
&-& {{3 \mp^2 \k1^4} \over \ds^4} \left[ F(s) - {{3 \sa} \over
\sqrt{s}}\as \right. \no \\
&-& \left. F(\k1^2) + {{3 \ka} \over \sqrt{\k1^2}}\ak\right],
\end{eqnarray}

\begin{eqnarray}
{I_2(z^2_1 z_2) \over m^2_K} &=& -{1 \over {6 \ds}} + {1 \over
\ds^2}\left[ -{{4 \mp^2}
\over 3} \right. \no \\
&-& \left. {{\sqrt{\k1^2} \ka} \over 3}\ak \right. \no \\
&+& \left. {{4 \mp^2} \over {3\sqrt{\k1^2}}}\ka \ak \right]
- {1 \over \ds^2}\left[-2 \mp^2 + {{2 \k1^2 \mp^2} \over {3 s}}
\right. \no \\
&-& \left. \sa {{(2 \k1^2 \mp^2 +\k1^2 s - 6 \mp^2 s)}
\over {3 s \sqrt{s}}}\as \right], \no \\
\end{eqnarray}

\begin{eqnarray}
{I_2(z^2_1 z^2_2) \over m^2_K} &=& -{1 \over {24 \ds}} -{1 \over
{12\ds^2}}(6 \mp^2 -
\k1^2)-{\mp^4 \over \ds^3}\left[F(s) - F(\k1^2)\right] \no \\
&-& {{\k1^4 \over \ds^3}}\left[-{\mp^2 \over {3s}} + {\ns \over {6
s \sqrt{s} \sa}}\as \right. \no \\
&+& \left. {\mp^2 \over {3\k1^2}} - {\nk \over {6 \k1^2 \sqrt{\k1^2}
\ka}}\ak \right] \no \\
&+& {{2 \mp^2 \k1^2} \over \ds^3}\left[-{\ka \over \sqrt{\k1^2}}\ak
\right. \no \\
&+& \left. {\sa \over \sqrt{s}}\as \right],
\end{eqnarray}

\begin{eqnarray}
{I_2(z^3_1 z_2) \over m^2_K} &=& -{1 \over {12 \ds}} + {1 \over
{2\ds^2}}\left[ -{{4 \mp^2}
\over 3} \phantom{\ak}\right. \no \\
&-& \left. {{\sqrt{\k1^2} \ka} \over 3}\ak \right. \no \\
&+& \left. {{4 \mp^2} \over {3\sqrt{\k1^2}}}\ka \ak \right]
-{1 \over {2\ds^2}}\left[-2 \mp^2 + {{2 \k1^2 \mp^2} \over {3 s}}
\right. \no \\
&-& \left. \sa {{(2 \k1^2 \mp^2 +\k1^2 s - 6 \mp^2 s)} \over {3 s
\sqrt{s}}}\as \right], \no \\
\end{eqnarray}

\begin{eqnarray}
I_3 m^2_K &=& -{13 \over 12}\mp^2 + {13 \over 144}\left(s +\k1^2\right)
- {\mp^4 \over {2 \ds}}\left[F(s) - F(\k1^2)\right] \no \\
&-& {\k1^4 \over {2 \ds}}\left[
- {\nk \over {6\k1^2}\sqrt{\k1^2} \ka} \ak \right. \no \\
&+& \left. {\ns \over {6 s\sqrt{s} \sa}} \as \right] \no \\
&+& {{\k1^2 \mp^2} \over {s-k^2_1}} \left[-{\ka \over
\sqrt{\k1^2}}\ak \right. \no \\
&+& \left. {\sa \over \sqrt{s}}\as\right] \no \\
&-& \sa {{\left(2 \k1^2 \mp^2 +\k1^2 s -10 \mp^2 s +s^2\right)} \over
{12 s \sqrt{s}}}\as, \no \\
\end{eqnarray}

\begin{eqnarray}
I_4 m^2_K = {{5 \k1^2} \over {18}} - {{4 \mp^2} \over 3} + {{(4 \mp^2 -
\k1^2) \ka} \over {3 \sqrt{\k1^2}}}\ak,
\end{eqnarray}

\begin{eqnarray}
\label{last}
I_5 = -{8 \over 9} + {{8 \mp^2} \over {3 \k1^2}} - {{2(4 \mp^2
-\k1^2) \ka} \over {3 \k1^2 \sqrt{\k1^2}}}\ak.
\end{eqnarray}

\n In the cases when $s > 4 \mp^2$ or $\k1^2 > 4 \mp^2$, we have to
perform the substitutions

\begin{eqnarray}
\as &\ri& -{1 \over {2i}}\left[\log\left({{1-\sqrt{1-4\mp^2/s}} \over
{1+\sqrt{1-4\mp^2/s}}}\right)+i\pi\right], \no \\
\ak &\ri& -{1 \over {2i}}\left[\log\left({{1-\sqrt{1-4\mp^2/\k1^2}} \over
{1+\sqrt{1-4\mp^2/\k1^2}}}\right)+i\pi\right],
\end{eqnarray}

\n and

\begin{eqnarray}
\sa &\ri& i\sqrt{s-4 m^2_{\pi}}, \no \\
\ka &\ri& i\sqrt{k^2_1-4 m^2_{\pi}},
\end{eqnarray}

\n respectively, in formulas (\ref{first})--(\ref{last}).

\vfill\eject

\end{document}